\documentstyle[12pt,epsf]{article}
\epsfverbosetrue
\def\figlabel#1{\xdef#1{\thefigure}}

\def\figalign#1#2#3#4#5#6{
\begin{figure}
\centerline{
\hbox to 2.5truein{\vtop{\hsize=2.5truein\epsfxsize=6cm
\centerline{\epsfbox{#1} }
\caption[]{#3}
\figlabel{#2} }}
\qquad\hbox to 2.5truein{\vtop{\hsize=2.5truein\epsfxsize=6cm
\centerline{\epsfbox{#4} }
\caption[]{#6}
\figlabel{#5} }} }
\end{figure} }

 \def\ex{{\hbox{\rm e}}}

\def\im{{\hbox{\rm Im}}}  \def\tr{{\hbox{\rm 
Tr}}}

\def\ie{{\em i.e.}}
\def\be{\begin{equation}}
\def\ee{\end{equation}}
\def\bea{\begin{eqnarray}}
\def\eea{\end{eqnarray}}

\def\im{{\hbox{\rm Im}}}

\def\ad{\hbox{\rm ad}}

\def\intl{\int\limits}
\def\ex{{\hbox{\rm e}}}  
\def\tr{{\hbox{\rm Tr}}}
\def\too{\longrightarrow}
\def\half{{1\over 2}} 

\def\to{\rightarrow}

\def\sqr#1#2{{\vcenter{\vbox{\hrule height.#2pt
  \hbox{\vrule width.#2pt height#1pt \kern#1pt
    \vrule width.#2pt}
  \hrule height.#2pt}}}}

\def\vocal{\langle {\cal O} \rangle}
\def\vocalv{\langle V \rangle}
\def\dalpha{{\dot\alpha}}

\def\ll{\left}
\def\rr{\right}

\font\cmss=cmss12 \font\cmsss=cmss12 at 8pt
\def\IR{\relax{\rm I\kern-.18em R}}

\def\IL{\relax{\rm I\kern-.18em L}}
\def\IH{\relax{\rm I\kern-.18em H}}
\def\IR{\relax{\rm I\kern-.18em R}}
\def\IC{\relax\hbox{$\inbar\kern-.3em{\rm C}$}}
\def\IZ{\relax\ifmmode\mathchoice
{\hbox{\cmss Z\kern-.4em Z}}{\hbox{\cmss Z\kern-.4em Z}}
{\lower.9pt\hbox{\cmsss Z\kern-.4em Z}}
{\lower1.2pt\hbox{\cmsss Z\kern-.4em Z}}\else{\cmss Z\kern-.4em
Z}\fi}
\def\IB{\relax{\rm I\kern-.18em B}}
\def\IC{{\relax\hbox{$\inbar\kern-.3em{\rm C}$}}}
\def\ID{\relax{\rm I\kern-.18em D}}
\def\IE{\relax{\rm I\kern-.18em E}}
\def\IF{\relax{\rm I\kern-.18em F}}
\def\IG{\relax\hbox{$\inbar\kern-.3em{\rm G}$}}
\def\IGa{\relax\hbox{${\rm I}\kern-.18em\Gamma$}}
\def\IH{\relax{\rm I\kern-.18em H}}
\def\II{\relax{\rm I\kern-.18em I}}
\def\IK{\relax{\rm I\kern-.18em K}}
\def\IP{\relax{\rm I\kern-.18em P}}
\def\IQ{\relax\hbox{$\inbar\kern-.3em{\rm Q}$}}

\def\hat{\widehat}

\def\cn {{\cal N}}

\begin{document}

\begin{titlepage}
\begin{flushright} { ~}  USC-FT-2/99\\
hep-th/9901161\\ 
\end{flushright}
\vspace*{20pt}
\bigskip
\begin{center}
{\Large Duality in the Context of }
\vskip3mm
{\Large Topological Quantum Field
Theory\footnote{Invited lecture delivered by J.M.F. Labastida at the
workshop on ``New Developments in Algebraic Topology" held at Faro on July
13-15, 1998.}}
\vskip 0.9truecm

\vspace{3pc}

{J. M. F. Labastida and  Carlos Lozano}

\vspace{1pc}

{\em  Departamento de F\'\i sica de Part\'\i culas,\\ Universidade de
Santiago de Compostela,\\ E-15706 Santiago de Compostela, Spain.\\}

\vspace{10pc}

{\large \bf Abstract}
\end{center} 

We present a summary of the progress made in the last few years on
topological quantum field theory in four dimensions. In particular, we
describe the role played by duality in the developments which led to the 
Seiberg-Witten invariants and their relation to the Donaldson invariants. In
addition, we analyze the fruitful framework that this connection has
originated. This analysis involves the study of topological quantum
field theories which contain twisted $\cn=2$ supersymmetric matter fields 
as well as theories obtained after twisting $\cn=4$ supersymmetry. In the 
latter case, we present some recent results including the generalization of 
the partition function of the Vafa-Witten theory for gauge group $SU(N)$ with
prime $N$.

\end{titlepage}



Topological quantum field theory (TQFT) in four dimensions
\cite{tqft}\cite{thompson}\cite{moore}\cite{laplata} has become a  very fruitful
link between physics and mathematics. On the one hand, the progress made in the
last years on the duality properties of $\cn=2$ supersymmetric Yang-Mills
theories has been applied to their topological counterparts to define new
invariants, the Seiberg-Witten invariants \cite{sw}, and to show their relation
with known invariants as the Donaldson invariants
\cite{donald}. On the other hand the topological nature of twisted $\cn=4$
supersymmetric theories has provided important tests of our ideas on duality
symmetry. Both sides  benefit from each other and certainly they will continue
to do so in the forthcoming years after the consequences of the recent AdS/CFT
conjecture \cite{malda}\cite{klebanov}\cite{wittenads} in this context start
being  explored -- see \cite{hull}\cite{gopaku} for some proposals in this 
direction.

Edward Witten inaugurated the field of topological quantum field theory in the
beginning of 1988 with his work on Donaldson theory from a quantum field theory
perspective \cite{tqft}. He formulated a twisted version of
$\cn=2$ supersymmetric gauge theory, now known as Donaldson-Witten theory, whose
observables were identified with the Donaldson invariants of  four-manifolds
\cite{donald}. His formulation was later reinterpreted \cite{jeffrey} from a
more geometrical point of view, in terms of a representative of the Thom class
of a vector bundle associated to certain moduli problem in the framework of the
Mathai-Quillen formalism \cite{mathai}. Twisted $\cn=2$ supersymmetric theories,
in general, are associated to certain moduli problems which, properly treated in
the context of the Mathai-Quillen formalism, lead to representatives of the Thom
class which become the exponential of the twisted actions on the field theory
side. Both pictures of Donaldson-Witten theory have been known for some time.
One important property of the resulting TQFT is that the vacuum expectation
values of its observables are independent of the coupling constant. This means
that these quantities could be computed in either the strong or the weak
coupling limit. The weak coupling limit analysis showed  the relation of the
observables of the theory to the Donaldson invariants. However, in such analysis
no new progress was made from the quantum field theory representation regarding
the calculation of these invariants. The difficult problems that one had to face
were similar to those in ordinary Donaldson theory. 

The strong-coupling effective theory of $\cn=2$ supersymmetric Yang-Mills theory
was obtained by  Seiberg and Witten in 1994 \cite{sw}. One would expect that the
twisted version of this effective theory would be related to the
Donaldson-Witten theory. Furthermore, since the  observables of a TQFT are
independent of the coupling constant, the weak coupling limit of the effective
theory should be exact, \ie, it would lead to Donaldson invariants. This is in
fact what turns out to be the case. The twisted effective theory could be
regarded as a TQFT {\it dual} to the original one. In addition, one could ask
for the {\it dual} moduli problem associated to this dual TQFT. It turns out
that in some of the most interesting situations ($b_2^+>1$) this moduli space is
an Abelian version of the moduli space of instantons modified by the presence of
chiral spinors. This space is known as the moduli space of Abelian monopoles
\cite{abm}. Being related to an Abelian gauge theory this space is simpler to
analyze than the moduli space of instantons. Furthermore,  for a large set of
four-manifolds (of {\sl simple type}), only particular classes of Abelian gauge
configurations (basic classes) contribute. For these classes the moduli space of
Abelian monopoles reduces to a finite set of points.

Donaldson-Witten theory has been generalized  after studying its coupling to
topological matter fields \cite{rocek}\cite{top}\cite{ans}. The resulting theory
can be regarded as a twisted form of $\cn=2$ supersymmetric Yang-Mills theory
coupled to hypermultiplets, or, in the context of the Mathai-Quillen formalism,
as the TQFT associated to the moduli space of non-Abelian monopoles
\cite{abmono}\cite{nabm}. Perturbative and non-perturbative methods have been
applied to this theory for the case of gauge group
$SU(2)$  and one hypermultiplet of matter in the fundamental representation
\cite{last}. In this case, again, it turns out that when 
$b_2^+>1$ the generalized Donaldson invariants can be written in terms of
Seiberg-Witten invariants. 

Recently, a general framework to analyze models with gauge group $SU(2)$, known
as integration over the $u$-plane
\cite{moorewitten}\cite{lns}\cite{mmone}, has been constructed. From this new
viewpoint, the presence of Seiberg-Witten invariants turns out  to be rather
general. They are believed to provide the only contributions to the invariants
for manifolds with
$b_2^+>1$. Generalizations to higher-rank gauge groups have been also studied
\cite{mooremari}.  In all these examples one finds relations among different
moduli spaces. In the context of TQFT in four
dimensions, duality relates moduli spaces: observables which are topological
invariants of a given four-manifold can be computed using information from two
different moduli spaces. The duality properties of the physical theory fix the
type of moduli spaces which are involved in each case.

Not all the theories obtained after twisting extended supersymmetric theories
fall into the duality pattern among moduli spaces described above. That is the
case for some of the twistings which originate from  $\cn=4$ supersymmetric
Yang-Mills theory, in particular for the twisting considered in
\cite{vafawitten} by Vafa and Witten. From the duality  point of view  discussed
in this paper this theory can be regarded as a self-dual theory in the sense
that it involves only one moduli space. In this case duality manifests itself in
a different form: it becomes an $SL(2,\IZ)$ symmetry which involves the coupling
constant and the dual gauge groups
\cite{vafawitten}.

To begin with the description of Donaldson-Witten theory we first
review some generalities  concerning $\cn=2$
supersymmetry in  four-dimensions. The global symmetry group of $\cn=2$
supersymmetry in
${\IR}^4$ is 
${\cal H}= SU(2)_L\otimes SU(2)_R \otimes SU(2)_I \otimes U(1)_{\cal R}$  where
${\cal K} = SU(2)_L \otimes SU(2)_R$ is the rotation group and 
$SU(2)_I \otimes U(1)_{\cal R}$ is the internal (chiral) symmetry  group.
The supercharges, $Q^i_\alpha$ and $\overline Q_{i\dot\alpha }$, which 
generate
$\cn=2$ supersymmetry, have the following transformations under ${\cal H}$:
\begin{equation} Q^i_\alpha \;\; \left(\frac{1}{2},0,\frac{1}{2}\right)^1,
\;\;\;\;\;\;\;\;\;\;\;
\overline Q_{i \dot\alpha } \;\;
 \left(0,\frac{1}{2},\frac{1}{2}\right)^{-1},
\label{ctreintatres}
\end{equation} where the superindex denotes the $U(1)_{\cal R}$ charge and
the numbers within parentheses label the representations under each of the
factors in 
$SU(2)_L\otimes SU(2)_R \otimes SU(2)_I$. The supercharges  
(\ref{ctreintatres}) satisfy: 
\begin{equation}
\{Q^i_\alpha, \overline Q_{j\dot\beta} \} = \delta^i_j 
P_{\alpha\dot\beta}.
\label{ctreintacuatro}
\end{equation}

The twist consists of considering as the rotation group the group,
${\cal K}' = SU(2)_L'\otimes SU(2)_R$, where $SU(2)_L'$ is the diagonal  
subgroup of $SU(2)_L\otimes SU(2)_I$. This implies that the isospin index
$i$ becomes a spinorial index $\alpha$: $Q^i_\alpha \rightarrow
Q^\beta_\alpha$ and
$\overline Q_{i\dot\beta } \rightarrow G_{\alpha\dot\beta}$.  The
trace of $Q^\beta_\alpha$ is chosen as the generator of a new scalar 
symmetry:
$Q = Q^\alpha_\alpha$. Under the new global symmetry group ${\cal H}'={\cal
K}'\otimes U(1)_{\cal R}$, the symmetry generators transform as:
\begin{equation} G_{\alpha\dot\beta} \;\;
\left(\frac{1}{2},\frac{1}{2}\right)^{-1},
\;\;\;\;\;\;\;\;\;\; Q_{(\alpha\beta)} \;\; (1,0)^1,
\;\;\;\;\;\;\;\;\;\; Q \;\; (0,0)^1.
\label{ctreintacinco}
\end{equation}
It is important to 
stress that as long as we stay on a flat space (or one with trivial
holonomy),  the twist is just a fancy way of considering the theory, for
in the end we  are not changing anything. However, the appearance of a
scalar symmetry  makes the procedure meaningful when we move to an
arbitrary four-manifold.  Once the scalar symmetry is found we must study
if it can be written as the transformation of some quantity under $Q$. 
If this is the case, the vacuum expectation value of $Q$-invariant
operators will be metric independent. The
$\cn=2$ supersymmetry algebra gives a  necessary condition for this to 
hold. Indeed, after the twisting, this algebra becomes:
\begin{equation}
\{Q^i_\alpha, \overline Q_{j\dot\beta} \} = \delta^i_j  P_{\alpha\dot\beta}
\longrightarrow
\{ Q , G_{\alpha\dot\beta} \} = P_{\alpha\dot\beta},
\label{ctreintaseis}
\end{equation} where $P_{\alpha\dot\beta}$ is the momentum operator of the
theory. Certainly (\ref{ctreintaseis}) is only a necessary condition for
the theory to be  topological. However, up to date, for all the 
supersymmetric  models whose twisting has been studied the relation on the
right hand side of (\ref{ctreintaseis}) has become valid for the whole
energy-momentum tensor. 

In ${\IR}^4$ the original and the twisted theories are equivalent. 
However, for arbitrary manifolds $X$ they are certainly different due to the
fact that  their energy-momentum tensors are not the same. The twisting
changes the spin  of the fields in the theory, and therefore their
couplings  to the metric on
$X$ become modified. This suggests an alternative way of  looking at the
twist. All that has to be done is: gauge the internal group
$SU(2)_I$, and identify the corresponding $SU(2)$ connection with the spin
connection on $X$. This process changes the spin connection and therefore
the energy-momentum tensor of the theory,  which in turn modifies the
couplings  to gravity of the different fields of the theory. This
alternative point of view to the twisting procedure has been 
reviewed in this context in
 \cite{baryon}. 

As mentioned above, the Donaldson-Witten theory can be constructed by twisting
the pure
$\cn$=2 supersymmetric Yang-Mills theory with gauge group $SU(2)$. This theory
contains a gauge field $A$, a pair of chiral spinors $\lambda_i$ and a complex
scalar field $B$. Under the twist, this field content is modified as follows:
\bea A_{\alpha\dalpha}\,\,\ll(\half,\half,0\rr)^0 &\too&
A_{\alpha\dalpha}\,\,\ll(\half,\half\rr)^0,\nonumber\\
\lambda_{\alpha i}\,\,\ll(\half,0,\half\rr)^{-1} &\too&
\chi_{\alpha\beta}\,\,(1,0)^{-1},\,\,\eta (0,0)^{-1},\nonumber\\
\bar\lambda^j_{\dalpha}\,\,\ll(0,\half,\half\rr)^1 &\too&
\psi_{\alpha\dalpha}\,\,\ll(\half,\half\rr)^1,\nonumber\\  B\,\,\,(0,0,0)^{-2}
&\too&
\lambda\,\,\,(0,0)^{-2},
\nonumber\\ B^{*}\,\,\,(0,0,0)^2 &\too&
\phi\,\,\,(0,0)^2.
\label{ctreintasiete}
\eea In the process of twisting, the $U(1)_{\cal R}$ symmetry becomes the  
$U(1)$-like symmetry associated to the ghost number of the topological
theory. The  ghost number anomaly is thus naturally related to the chiral
anomaly of
$U(1)_{\cal R}$.
The twisted action has the form:
\bea
 && \int_X d^4
x\,\sqrt{g}\tr\Bigg({F^+}^2-i\chi^{\mu\nu} D_\mu \psi_\nu +i\eta D_\mu
\psi^\mu +\frac{1}{4} \phi \{ \chi_{\mu\nu},\chi^{\mu\nu} \}
+\frac{i}{4} \lambda \{\psi_\mu,\psi^\mu\}
-\lambda D_\mu D^\mu\phi  \nonumber  
\\ &&\,\,\,\,\,\,\,\,\,\,\,\,\,\,\,\,\,\,\,\,\,\,\,\,\,\,\,\,\,\,\,\,\,\,\,
\,\,\,\,\,\,\,\,\,\,\,\,\, +
\frac{i}{2}\phi\{\eta,\eta\}
+\frac{1}{8}[\lambda,\phi]^2\Bigg).
\label{ctreinta}
\eea
It is invariant under the transformations generated by $Q$ which from now
on will be denoted as $\delta$-transformations: 
\begin{eqnarray}
\delta A_\mu &= \psi_\mu,  \mbox{\hskip2cm} \delta \chi_{\mu\nu} &=  
G_{\mu\nu},
\nonumber \\ \delta \psi \,&= d_A\phi, \mbox{\hskip2.2cm}
\delta\eta & = i [\lambda,\phi],  \nonumber \\
\delta \phi \,&= 0, \mbox{\hskip2.6cm}
\delta\lambda &= \eta.
\mbox{\hskip2cm} 
\label{cveintenueve}
\end{eqnarray} 
In these transformations, $\delta^2$ is a gauge transformation with gauge
parameter $\phi$. Observables are therefore related to the  ${\cal
G}$-equivariant cohomology of $\delta$ (that is, the cohomology of $\delta$ 
restricted to gauge invariant operators). Of course, auxiliary fields can be
introduced so that the action (\ref{ctreinta}) is
$\delta$-exact
\cite{laplata}.

To construct the observables of the theory we begin by pointing out
that for each independent Casimir of the gauge group
$G$  it is possible to construct a highest-ghost-number operator $W_0$, from
which lower ghost-number operators $W_i$ can be defined recursively through the
descent equations $\delta W_i = d W_{i-1}$. For example, for  the quadratic
Casimir this operator is:
 \be W_0 = \frac{1}{8\pi^2} \tr\, (\phi^2), 
\label{raquela}
\ee  and it generates the following family of operators:
\be W_1 = \frac{1}{4\pi^2}\tr\,(\phi \psi), \,\,\,\,\,\,\,\,\,\,\,\,\,\, W_2
=
\frac{1}{4\pi^2} \tr\,\ll(\frac{1}{2}\psi\wedge\psi+\phi\wedge F\rr),
 \,\,\,\,\,\,\,\,\,\,\,\,\,\, W_3 = \frac{1}{4\pi^2} \tr\,(\psi\wedge F).
\label{raquel}
\ee  From them one defines the following observables:
\begin{equation} {\cal O}^{(k)} = \int_{\gamma_k} W_k,
\label{sonia}
\end{equation} where $\gamma_k \in H_k(X)$. The descent equations imply
that they are
$\delta$-invariant and that they only depend on the homology class
 of $\gamma_k$. According to  (\ref{ctreintasiete}), the ghost numbers of these
operators are $U({\cal O}^{(k)})=4-k$.

The functional integral corresponding to the topological invariants  of 
the theory has the form:
\begin{equation}
\ll\langle {\cal O}^{(k_1)} {\cal O}^{(k_2)} \cdots {\cal O}^{(k_p)} \rr\rangle =
\int  {\cal O}^{(k_1)} {\cal O}^{(k_2)} \cdots {\cal O}^{(k_p)} \exp
({-{S}/{g^2}}),
\label{clara}
\end{equation} where the integration has to be understood on the space of
field  configurations modulo gauge transformations, and $g$ is a coupling
constant. Standard arguments show that due to the
$\delta$-exactness of the action $S$, the  quantities obtained in
(\ref{clara}) are independent of $g$. This implies that  the observables
of the theory can be obtained either in the limit
$g\rightarrow 0$, where perturbative methods apply, or in the limit
$g\rightarrow \infty$, where one is forced to consider a  non-perturbative
approach. 

Let us consider first the theory in the weak coupling limit $g\to 0$. The
previous argument affirms that the semiclassical approximation is exact. In
the weak coupling limit the contributions  to the functional integral are
dominated by the bosonic field configurations  which minimize
$S$. These turn out to be given by the equations:
\be  F^+=0,\qquad D_\mu D^\mu\phi=0.
\label{ctreintaocho}
\ee 
Let us assume that in the situation under consideration there are only
irreducible connections  (this is true in the case $b^+_2={\rm
dim}\,H^{2,+}(X)>1$). In this case  the contributions from the bosonic  
part of the action are given entirely by  the solutions of the equation
$F^+=0$,
\ie, by instanton configurations. Since the connection is irreducible, there
are no non-trivial  solutions to the second equation in
(\ref{ctreintaocho}).

The zero modes of the field $\psi$ come from the  solutions to the
equations
\begin{equation} (D_\mu\psi_\nu)^+=0, \;\;\;\;\;\;\;\; D_\mu\psi^\mu=0,
\label{ctreintanueve}
\end{equation} 
which define precisely  the tangent space
to the space of instanton  configurations. The number of independent
solutions of these equations determine the  dimension of the instanton
moduli space ${\cal M}_{\rm ASD}$. For
$SU(2)$ this dimension is $d_{{\cal M}_{\rm ASD}} = 8k-3(\chi+\sigma)/2$,
where $k$ is the instanton number, while $\chi$ and $\sigma$ are the Euler
characteristic and the signature of the manifold
$X$, respectively. 

The fundamental contribution to the functional integral (\ref{clara})  is
given by the elements of ${\cal M}_{\rm ASD}$ and by the zero-modes of the
solutions to (\ref{ctreintanueve}). Once these have been obtained they
must be introduced in the action and an expansion up to quadratic terms in
non-zero modes  must be performed. The fields $\phi$ and $\lambda$ are
integrated out originating a contribution \cite{tqft} which is equivalent
to the replacement of the field $\phi$ in the operators ${\cal O}^{(k)}$
by
\be
\phi^a \longrightarrow\int d^4y
\sqrt{g}\,G^{ab}(x,y)[\psi_\mu(y),\psi^\mu(y)]^b,
\label{smock}
\ee where $G^{ab}(x,y)$ is the inverse of the Laplace operator,
\be D_\mu D^\mu  G^{ab}(x,y)=\delta^{ab}\delta^{(4)}(x-y).
\label{propa}
\ee 
These are the only relevant terms in the limit
$g\rightarrow  0$. The resulting Gaussian integrations then must be
performed. Due to the  presence of the $\delta$ symmetry these come in
quotients whose value is $\pm 1$. The functional integral (\ref{clara})
takes the form: 
\begin{eqnarray} && \ll\langle {\cal O}^{(k_1)} {\cal O}^{(k_2)} \cdots {\cal
O}^{(k_p)} \rr\rangle = \nonumber \\ && \int_{{\cal M}_{\rm ASD}} da_1\cdots
da_{d_{{\cal M}_{\rm ASD}}} d\psi_1
\cdots d\psi_{d_{{\cal  M}_{\rm ASD}}} {\cal O}^{(k_1)} {\cal O}^{(k_2)}
\cdots {\cal O}^{(k_p)}  (-1)^{\nu(a_1,\dots,a_{d_{{\cal M}_{\rm
ASD}}})},
\label{clarados}
\end{eqnarray} where $\nu(a_1,\dots,a_{d_{{\cal M}_{\rm ASD}}})=0,1$. The
integration over the  odd modes leads to a selection rule for the product
of observables. This selection rule  is better expressed making use of the
ghost numbers of the fields. For the operators  in (\ref{sonia}) the 
selection rule can be written as $d_{{\cal M}_{\rm ASD}}=\sum_{i=1}^p
U({\cal O}^{(k_i)})=\sum_{i=1}^p (4-k_i)$.

In the case in which $d_{{\cal M}_{\rm ASD}}=0$, the only observable is
the partition function, which takes the form:
\begin{equation}
\langle 1 \rangle = \sum_i (-1)^{\nu_i},
\label{vanessa}
\end{equation} where the sum is over isolated instantons, and $\nu_i=0,
1$. In  general, the integration over the zero-modes in  (\ref{clarados})
leads to an antisymmetrization in such a way that one ends up with the
integration of  a
$d_{{\cal M}_{\rm ASD}}$-form on ${\cal M}_{\rm ASD}$. The resulting real
number is a  topological invariant of the four-manifold $X$. Notice that in the
process a map 
\be H_k(X) \longrightarrow  H^k({\cal M}_{\rm ASD}),
\label{prodi}
\ee has been constructed. Hence, the correlation functions of the topological
theory give polynomials in
$H_{k_1}({\cal M}_{\rm ASD}) \times H_{k_2}({\cal M}_{\rm ASD}) \times \cdots
\times  H_{k_p}({\cal M}_{\rm ASD})$, which are precisely the Donaldson
polynomial invariants of $X$.

The vacuum expectation values
(\ref{clarados})  can be collected in a very convenient way by introducing a
generating function. Let us assume that we consider only smooth, compact,
oriented four-manifolds which are simply connected. In this case only 
$W_0$ in (\ref{raquela}) and $W_2$ in (\ref{raquel}) are
relevant operators. Given a basis 
$\{\Sigma_a\}_{a=1,\dots,b_2(X)}$ of $H_2(X)$, we define, following
\cite{abm}\cite{moorewitten}\cite{wijmp}, the generating function
\be
F(\lambda,\alpha_1,\alpha_2,\dots) = \ll\langle 
\ex^{\sum_{a}\alpha_a I(\Sigma_a)+\lambda{\cal O}}
\rr\rangle,
\label{generador}
\ee
where ${\cal O}=W_0$ as in (\ref{raquela}), and $I(\Sigma_a)=\int_{\Sigma_a}
W_2$ as in (\ref{sonia}). In (\ref{generador}) the quantities $\alpha_a$,
$a=1,\dots,b_2(X)$ and $\lambda$ are constant parameters. The expectation values
(\ref{clarados}) can be easily extracted from  the power expansion of
(\ref{generador}).

In the weak coupling limit one finds for the function
$F(\lambda,\alpha_1,\alpha_2,\dots)$ in (\ref{generador}) an expression similar
to (\ref{clarados}). This expression does lead to the computation difficulties
inherent in the  Donaldson theory. In the weak coupling limit one proves that
the  Donaldson theory has a quantum field theory interpretation but this
interpretation does not provide new insights to compute the Donaldson invariants.
Nevertheless, the field theory connection is very important since in this theory
the strong and weak coupling limits are exact, and therefore the door is
open to find a strong coupling description which could lead to a 
new, simpler representation for the Donaldson invariants. The strong coupling
realization of the Donaldson-Witten theory was found by Witten \cite{abm} after
using the results on the strong coupling behavior of $\cn=2$ supersymmetric
gauge theories which he and N. Seiberg \cite{sw} had discovered. The key
ingredient used by Witten was to assume that the strong coupling limit of
Donaldson-Witten theory is equivalent to the ``sum" over the twisted
effective low energy descriptions of the corresponding $\cn=2$ physical
theory. This ``sum" is  not entirely a sum, as in general it has a part which
contains a continuous integral. The ``sum" is now known as integration
over the
$u$-plane after the work by Moore and Witten \cite{moorewitten}. The
reasons for this will become clear below. Witten's assumption in \cite{abm}
can be simply stated as saying that the weak-strong coupling limit and the
twist commute. In other words, to study the strong coupling limit of the
topological theory, first one untwists, then one works out the strong
coupling limit of the physical theory and, finally, one twists back. 

In order to implement the duality picture among moduli spaces we need to
know two important pieces of information: the low energy description of
$\cn=2$ extended supersymmetric Yang-Mills theories and the map of the
topological observables (\ref{sonia}) into their strong coupling
counterparts. The first of these issues was addressed by Seiberg and Witten
in 1994 \cite{sw} and its basic structure is by now well-known in rather
general situations. The second  was discussed in several works
\cite{abm}\cite{last}\cite{wijmp} but only recently it has been systematized
using the canonical solution to the descent equations used for the first time,
though in a different TQFT, in \cite{llatas}.

 From the work by Seiberg and Witten \cite{sw} follows that at low 
energies
$\cn=2$ supersymmetric Yang-Mills theories behave as Abelian gauge 
theories. For the case of gauge group $SU(2)$, which will be the case 
considered in this discussion, the effective low energy theory is
parametrized  by a complex variable $u$ which labels the vacuum structure
of the theory.  At each value of
$u$ the effective theory is an $\cn=2$ supersymmetric Abelian gauge theory
coupled to $\cn=2$ supersymmetric matter fields. One of the most salient
features of the low-energy description is that there are points in the
complex $u$-plane where some matter fields become massless. These  points
are singular points of the effective theory and they are located at
$u=\pm 1$.  At $u=1$ the effective theory consists of an  
$\cn=2$ supersymmetric Abelian gauge theory coupled to a massless monopole,  
while at $u=-1$ it is coupled to a dyon.  The effective theories
at each singular point are related by a chiral $\IZ_2$ symmetry which exists on
the
$u$-plane. This symmetry relates the behavior of the  theory around one
singularity to its behavior around the other. 

One of the most important
features of $\cn=2$ supersymmetric Yang-Mills theory is that its lagrangian
can be  written in terms of a single holomorphic function, the prepotential 
${\cal F}$.  This prepotential is holomorphic in the sense that it depends
only on an  $\cn=2$ chiral superfield $\Psi$ which defines the theory, and
not on its complex  conjugate.  The microscopic theory is defined by a
classical quadratic prepotential:
\be {\cal F}_{{{\rm cl}}}({\Psi})=\half \tau_{{{\rm cl}}}  {\Psi}^2,\qquad 
\tau_{\rm cl}={\theta_{\rm bare}\over2\pi}+{4\pi  i\over g^2_{\rm bare}}.
\label{ccincuentatres}
\ee In terms of this prepotential the lagrangian is given by the following
expression in $\cn=1$ superspace: 
\be {\cal L}={1\over{4\pi}}{\hbox{\rm Im}}\tr\left[\int d^4\theta  
{\partial{\cal F}(A)\over\partial A}\bar A+\int d^2\theta\half
{\partial^2{\cal F}(A)\over\partial A^2}W^\alpha W_\alpha\right],
\label{ccincuentados}
\ee where $A$ is a chiral $\cn=1$ superfield containing the fields  
$(\phi,\psi)$, and $W$ is a constrained chiral spinor superfield
containing the  non-Abelian gauge field and its $\cn=1$ superpartner
$(A_\mu,\lambda)$. All the  fields take values in the adjoint
representation of the gauge group, which we take to be $SU(2)$. The
potential term for the complex scalar $\phi$ is: 
\be V(\phi)=\tr\left([\phi,\phi^{\dag}]^2\right).
\label{ccincuentacuatro}
\ee The minimum of this potential is attained at field configurations of
the form $\phi=\half a\sigma^3$, which define the classical moduli space
of  vacua. A convenient gauge invariant parametrization of the vacua is
given by  
$u=\tr\phi^2$, which equals $\half a^2$ semiclassically.  For $u\not =0$,
$SU(2)$ is spontaneously broken to $U(1)$. The spectrum of the theory
splits up  into two massive $\cn=2$ vector multiplets, which accommodate the
massive
$W^{\pm}$ bosons together with their superpartners, and an $\cn=2$ Abelian
multiplet  which accommodates the $\cn=2$ photon together with its
superpartners.  For
$u=0$, the full $SU(2)$ symmetry is (classically) restored.

To study the quantum vacua Seiberg and Witten analyzed the structure of
the low energy  theory, whose effective lagrangian up to two derivatives
is given, after integrating  out the massive modes, by an expression like
(\ref{ccincuentados}) but with a  new effective prepotential depending
only on an Abelian multiplet. The  result of their analysis has some
important features. First of all, it turns out that at the quantum level the
$SU(2)$ symmetry is never restored, \ie,  the  theory stays in the Coulomb
phase throughout the
$u$-plane. The moduli space of vacua ($u$-plane) is a complex 
one-dimensional K\"ahler manifold in which the prepotential ${\cal F}$ has
singularities at the points $u=\pm 1$.
These singularities correspond to the presence of a massless monopole
(at  $u=1$) and a massless dyon (at $u=-1$).
Near each of the singularities, the complete non-singular effective action
should  include together with the $\cn=2$ Abelian vector multiplet, a massless
monopole or a dyon hypermultiplet.

 As we did in the
perturbative approach, we will consider the theory  on  manifolds $X$ with
$b^+_2>1$. In the limit $g\to 0$  one has to take into account the
classical moduli space. Since for $b^+_2>1$  there are not Abelian
instantons the only contribution comes from $u=0$ and one  has to go
through the analysis carried out in our discussion of the perturbative
approach. As described there, one is led to the standard approach to 
Donaldson invariants via integration over the moduli space of non-Abelian 
instantons. On the other hand, in the limit  $g\to\infty$, since the
supersymmetric theory is  asymptotically free, we are in the infrared regime,
and the contributions come from  the quantum moduli space.  In the case under
consideration ($b^+_2>1$)  there are no Abelian instantons. Since the
Abelian gauge field is the only massless field away from the
singularities, the only contributions come from the singular points,
$u=\pm 1$, where there are additional massless fields. Near each of
these points, $\cn=2$ supersymmetry dictates the form of the  weakly coupled
effective theory. Since the observables of the twisted  theory are
independent of the coupling constant, one expects that Donaldson 
invariants can be expressed in terms of vevs of some operators in the
twisted  effective theories around each singular point. 

The theory around the monopole singularity is an $\cn=2$ supersymmetric
Abelian gauge theory coupled to a massless hypermultiplet. This theory has
a twisted version which has been constructed in  \cite{rocek}\cite{top} from the
point of view of twisting
$\cn=2$ supersymmetry, and in  \cite{abmono} using the Mathai-Quillen 
formalism. It has been addressed in other works \cite{ans}\cite{gian}. The
structure of this theory is similar to that of the Donaldson-Witten 
theory. The resulting action is $\delta$-exact and therefore one can study
the  theory in the weak coupling limit, which, as the theory is Abelian,
corresponds to the low  energy limit.

Let us describe the structure of the twisted $\cn=2$ supersymmetric  Abelian
gauge theory coupled to a twisted hypermultiplet. We will assume that the
four-dimensional manifold $X$ is a spin manifold. The analysis naturally
extends to the case of manifolds which are not spin as shown in
 \cite{abm}. A hypermultiplet is built out of two chiral
$\cn=1$ superfields, $Q$ and  $\tilde Q$,
\bea && Q(q^1,\psi_{q\alpha}),\qquad
Q^{\dag}(q^{\dag}_1,\bar\psi_{q\dalpha}),\nonumber\\ && \tilde
Q(q^{\dag}_2,\psi_{\tilde q\alpha}),\qquad \tilde Q^{\dag}  
(q^2,\bar\psi_{\tilde q\dalpha}).
\label{ccincuentacinco}
\eea 
After the twisting these fields become:
\bea q^i\,\,\ll(0,0,\half\rr)^0 &\too&
M^{\alpha}\,\,\ll(\half,0\rr)^0,\nonumber\\
\psi_{q\alpha }\,\,\ll(\half,0,0\rr)^{1} &\too&
\mu_{\alpha}\,\,\ll(\half,0\rr)^{1},\nonumber\\
\bar\psi_{\tilde q\dalpha}\,\,\ll(0,\half,0\rr)^{-1} &\too&
\nu_{\dalpha}\,\,\ll(0,\half\rr)^{-1},\nonumber\\ 
q^{\dag}_i\,\,\ll(0,0,\half\rr)^{0}&\too&
\overline M_\alpha\,\,\ll(\half,0\rr)^{0},\nonumber\\
\bar\psi_{q\dalpha}\,\,\ll(0,\half,0\rr)^{-1} &\too&
\bar\nu_\dalpha\,\,\ll(0,\half\rr)^{-1},\nonumber\\
\psi_{\tilde q\alpha}\,\,\ll(\half,0,0\rr)^{1} &\too&
\bar\mu_{\alpha}\,\,\ll(\half,0\rr)^1.
\label{ccincuentaseis}
\eea

The twisted fields  $M_\alpha,
\mu_\alpha$, and $\nu_\dalpha$ belong, respectively, to 
$\Gamma(S^{+}\otimes L)$ and  $\Gamma(S^{-}\otimes L)$, where
$S^{\pm}$ are the positive/negative chirality spin bundles and $L$ is a  
complex line bundle. The action of the twisted Abelian effective theory can
be found in  \cite{laplata}\cite{abmono}. In the weak-coupling limit, the main
contribution to the functional integral comes from a bosonic configuration given
by the solutions to the equations:
\begin{equation} F_{\alpha\beta}^+ + \frac{i}{2} \overline M_{(\alpha}
M_{\beta)} =0,
\;\;\;\;\; D_{\alpha\dalpha} M^\alpha = 0.
\label{francesca}
\end{equation}
 These equations  are known as monopole equations \cite{abm}. The tangent
space to the moduli space, ${\cal M}_{\rm AM}$, defined by  these
equations is given by the linearization of (\ref{francesca}), which 
happen to be the field equations:
\bea &&(d\psi)^+_{\alpha\beta}+{i\over2}(\overline  
M_{(\alpha}\mu_{\beta)}+\bar\mu _{(\alpha}M_{\beta)})=0,\nonumber\\
&&D_{\alpha\dalpha}\mu^\alpha+i\psi_{\alpha\dalpha}M^\alpha=0.
\label{ccincuentanueve}
\eea The dimension of the moduli space can be calculated from  
(\ref{ccincuentanueve}) by means of an index theorem, and turns out to be
\cite{abm}, 
\be  d_{{\cal M}_{\rm AM}}=\bigl(c_1(L)\bigr)^2 -\frac{2\chi+3\sigma}{4}.
\label{csesenta}
\ee The only contributions to the partition function come from $d_{{\cal  
M}_{\rm AM}}=0$ (isolated monopoles). Introducing  the shorthand notation,
$x=-2c_1(L)$,  we have: 
\be  d_{{\cal M}_{\rm AM}}=0 \Leftrightarrow x^2=2\chi+3\sigma.
\label{csesentauno}
\ee 
As in the Donaldson-Witten theory, the integration over the  
quantum fluctuations around the background  (\ref{francesca}) gives an  
alternating sum over the different monopole solutions for a given  class
$x$:
\be  n_x=\sum_i \epsilon_{i,x},\qquad \epsilon_{i,x}=\pm 1.
\label{csesentados}
\ee The $n_x$ are the partition functions of the twisted Abelian theory
for a fixed class $x$ (compare to (\ref{vanessa})). Those classes such that
(\ref{csesentauno}) holds and
$n_x\ne 0$ are  called {\it basic classes}. The quantities $n_x$ turn out
to constitute a new set of topological invariants for four-manifolds known
as Seiberg-Witten invariants. Other observables different than the $n_x$  
could be studied in the theory of Abelian monopoles. We have
restricted our attention only to the
partition function of this theory because it turns out that the Seiberg-Witten
invariants,
$n_x$, are the fundamental quantities that enter in the computation of the
generating function (\ref{generador}) for the case of manifolds of simple type.

To compute the partition function of the full
theory we must sum over classes and take into account the contribution from
each of the singularities on the $u$-plane. Instead of computing the
partition function we will concentrate our attention on the more general
generating function (\ref{generador}).
In computing this we must address the question of what  is
the form of the observables of Donaldson-Witten theory in terms of 
operators of the effective Abelian theory. To answer this question we will
use the expansion of the observables in the untwisted, physical theory,
together  with the canonical solution to the descent equations in the
topological Abelian theory. We follow the argument presented in 
\cite{laplata}\cite{moorewitten}.  Near the monopole singularity, the $u$
variable has  the   expansion \cite{sw}:
\be u(a_D)= 1 + \ll( {du \over da_D} \rr)_{0}a_D + {\rm  higher
\,\  order}, 
\label{expan}
\ee where $(du / da_D)_{0}= -2i $, while ``higher order" stands for
operators of  higher dimensions in the expansion. The field $a_D$
corresponds to the  field  $\phi_D$ of the topological Abelian theory
\cite{abmono}, while the  gauge-invariant parameter $u$ corresponds to the
observable  (\ref{raquela}). Therefore, the map among operators between the
microscopic and the macroscopic descriptions must be such that ${\cal O}$
becomes $u$ up to a  factor:
\be
{\cal O} \rightarrow \langle {\cal O} \rangle u,
\label{mapa1}
\ee
where $\vocal$ is a constant parameter. Once the operator ${\cal O}$ has
been identified one can use the canonical solution to the descent equations
of the macroscopic description to obtain the operator which must correspond
to $I(\Sigma)$ in (\ref{generador}). Actually we will concentrate on
$I(v)=\sum_a \alpha_a I(\Sigma_a)$ and we will think of $v$ as the formal
sum $v=\sum_a \alpha_a\Sigma_a$.  The starting point is the operator $u$
and the procedure is described in \cite{moorewitten}. One finds:
\be
I(v) \rightarrow
-{i \vocalv \over \pi \sqrt{2} } \int_v \ll(
{1\over 32} {d^2 u \over d a_D^2} \psi \wedge \psi -
{\sqrt{2}\over 4} {du \over da_D}(F_{D-} + G_+) \rr),
\label{mapa2}
\ee
where $F_{D-}$ is the anti-self-dual part of the Abelian field strength
$F_D$ and
$\psi$ and $G_-$ are the Abelian versions of the anticommuting vector and
the auxiliary field, respectively, in (\ref{ctreintasiete}). As in the case
of (\ref{mapa1}), the quantity $\vocalv$ is a constant parameter. Both
$\vocal$ and $\vocalv$ can be reabsorbed by  a rescaling of the parameters
$\alpha_a$ and $\lambda$ in (\ref{generador}). They will be fixed later
comparing the prediction from the TQFT to known
mathematical results for a given manifold assuming that the parameters
$\alpha_a$ and $\lambda$ in (\ref{generador}) can be identified with the
ones used in the referred mathematical results.

Once the mapping has been established, one must study the vacuum expectation
value (\ref{generador}) in the effective Abelian theory near the
singularity at $u=1$. For this theory the operator associated to ${\cal O}$ in
(\ref{mapa1}) becomes a $c$-number ($u=1$) and we can  make the
replacement:
\be
{\cal O} \rightarrow \vocal.
\label{remapa1}
\ee
The contribution from the operator associated to $I(v)$ is more
subtle. First of all, for simply connected manifolds one can ignore the
$\psi^2$ terms. In the part containing  $F_{D-}$ in (\ref{mapa2}) one can
replace 
$F_{D-}$ by $F_D$ since the difference is $Q$-exact. The integration of the
term containing an auxiliary field $G_+$  can be easily done and it turns
out that $\exp(I(v))$ is then mapped to
\cite{moorewitten}:
\be
\exp(I(v)) \rightarrow
\exp\Bigg( -{i \vocalv \over 4\pi}\int_v \Big({du \over da_D}F_D\Big)
+\vocalv^2 {(du/da_D)^2 \over 8\pi \im\tau} \,\, v_+ \cdot v_+ \Bigg),
\label{remapa2}
\ee
where $v_+ \cdot v_+$ denotes the intersection pairing on $H_2(X)$. 

The term $v_+ \cdot v_+$ in (\ref{remapa2}) is a first sign of the
presence of a contact term in the operator associated to $\exp(I(v))$
in the macroscopic description. On general grounds one expects that if a
$I(v)$ is mapped to $\tilde I(v)$, the product 
$I(\Sigma_1)\,I(\Sigma_2) \cdots I(\Sigma_n) $ is not mapped to the product
$\tilde I(\Sigma_1)\,\tilde I(\Sigma_2) \cdots \tilde I(\Sigma_n) $.
Rather, one would expect contributions at the intersections of the surfaces
$\Sigma_a$. Thus, it is natural to expect a term containing
$v\cdot v = v^2$ in the operator associated to $\exp(I(v))$. The
explicit form of this term has been worked out in \cite{moorewitten} using
arguments based on the modular invariance of the low energy description.
It turns out that the correct mapping is:
\be
\exp(I(v)) \rightarrow
\exp\Bigg( -{i \vocalv \over 4\pi}\int_v \Big({du \over da_D}F_D\Big)
+\vocalv^2 v^2 T(u) \Bigg),
\label{remapa3}
\ee
where
\be
T(u) = - {1\over 24}\Bigg( E_2(\tau)\ll({du\over da_D}\rr)^2 - 8 u\Bigg),
\label{esisei}
\ee
with $E_2(\tau)$ the Eisenstein series of weight $2$. In the limit $u\rightarrow
1$, one finds
$du/ da = -2 i$,  $T(1)=1/2$, and, therefore,  the operator
$\exp(I(v))$ becomes the $c$-number:
\be
\exp(I(v)) \rightarrow
\exp\Bigg( \vocalv \,\, x \cdot v + {\vocalv^2\over 2}v^2\Bigg),
\label{remapa4}
\ee
where $x= -  c_1(L^2) = -F_D/2\pi$ is the class which entered
(\ref{csesentados}).

We are now in a position to evaluate the correlation function
(\ref{generador}) at the monopole singularity. To do this we must take into
account (\ref{mapa1}) and (\ref{remapa3}), integrate over the space of
monopole solutions ${\cal M}_{\rm AM}$, and then take the limit
$a_D\rightarrow 0$. The factors (\ref{mapa1}) and (\ref{remapa3}) can be
accompanied by some other terms that could be present due to the fact that
the form of the twisted theory is not unique. Terms involving the $F_D\wedge
F_D$ or the Euler characteristic $\chi$ and the signature $\sigma$ of the
manifold $X$ could be added to the Lagrangian. The most general form of the
possible terms present has been analyzed in \cite{moorewitten}. Their
concrete form can be computed using the wall crossing techniques described
there. These terms are now known and one possesses an explicit expression
for the vacuum expectation value (\ref{generador}) which involves the
integration over the moduli spaces of monopole solutions and the
consideration of the limit $a_D\rightarrow 0$. The solution notably
simplifies if one assumes that the only contributions are the ones
corresponding to $d_{{\cal M}_{\rm AM}}=0$. Manifolds which satisfy this
condition are called of simple type.
In this situation the calculation simplifies because then one must consider
the limit $a_D\rightarrow 0$ of the operators (\ref{mapa1}) and
(\ref{remapa3}) which just corresponds to (\ref{remapa1}) and
(\ref{remapa4}). The integration over the zero-dimensional moduli space
leads to the Seiberg Witten invariants (\ref{csesentados}), and the
global coefficient can be easily obtained considering the limit
$a_D\rightarrow 0$ of the extra terms fixed imposing wall crossing
conditions. 

After summing over classes $x$, the contribution from the point
corresponding to the monopole singularity becomes:
\be 
C_1 \hbox{\rm exp}\ll({\vocalv^2\over 2} v^2 +\lambda \langle {\cal O}
\rangle\rr)   \sum_{x} n_{x} {\rm e}^  {\langle V \rangle v \cdot x},
\label{singu}
\ee 
where $C_1$ is a constant which turns out to be:
\be
C_1= 2^{1+{7\chi\over 4} + {11\sigma\over 4} }.
\label{elfactor}
\ee

Next, we must work out the contribution from the dyon
singularity at   
$u=-1$.  This contribution is related to the one from
$u=1$ by a  
$\IZ_2$ transformation, which is the anomaly-free symmetry on the
$u$-plane which remains after the breaking of the  chiral symmetry
$U(1)_{\cal R}$. Let us  begin by  recalling the transformations of the
fields entering the observables  under the $U(1)_{\cal R}$-transformations:
\bea  &&\psi^1_\alpha\too \ex^{-i\varphi}\psi^1_\alpha,\nonumber\\
&&\psi^2_\alpha\too \ex^{-i\varphi}\psi^2_\alpha,\nonumber\\ &&B\,\too
\ex^{-2i\varphi}B.\nonumber
\label{csetentasiete}
\eea Instanton effects break this symmetry down to ${\IZ}_8$ ($4N_c-2N_f$
in the general case of $SU(N_c)$ gauge group with $N_f$ hypermultiplets
in  the fundamental representation). Under this anomaly-free ${\IZ}_8$, 
\be B\too \ex^{-2i({2\pi\over8})}B= \ex^{-{i\pi\over2}}B,
\label{csetentaocho}
\ee and therefore,
\be  u=\tr(B^2)\too\ex^{-i\pi}u=-u,
\label{csetentanueve}
\ee which gives a ${\IZ}_2$ symmetry on the $u$-plane. This ${\IZ}_2$ 
symmetry relates the contributions to the vevs from $u=1$ to those  from
$u=-1$. Under the $\IZ_8$ symmetry, the observables 
 transform as follows:
\bea
 I(\Sigma_a)=\frac{1}{4\pi^2}\intl_{\Sigma_a}\tr\left(\phi
F+\half\psi\wedge\psi\right)&\too&
\ex^{-{i\pi\over2}}I(\Sigma_a)=-iI(\Sigma_a),\nonumber\\  {\cal
O}=\frac{1}{8\pi^2}\tr(\phi)^2&\too& \ex^{-i\pi}{\cal O}=-{\cal
O},
\label{cochenta}
\eea 
hence, using (\ref{singu}) one finds:
\bea u=1,\qquad&& C_1\exp{\displaystyle{\left({\vocalv^2\over 2}
v^2+\lambda\langle{\cal O}\rangle +
\langle V\rangle v \cdot x \right)}},\nonumber\\ u=-1,\qquad&&
C_2\exp{\displaystyle{\left(-{\vocalv^2\over 2} v^2-\lambda\langle{\cal
O}\rangle  -i\langle V\rangle v \cdot x \right)}}.
\label{cochentauno}
\eea

The quantities $C_2$ and $C_1$ are  related because on a curved 
background the ${\IZ}_8$ transformation, while being preserved by gauge 
instantons, picks anomalous contributions from the measure due to
gravitational  anomalies. The contribution is of the form
$\exp{{i\pi\over2}\Delta}$, where
$\Delta= {\chi+\sigma\over4}$. Notice that for a basic
class $x$, ${\rm dim}\,{\cal M}_{\rm AM}=0$, and therefore, from
(\ref{csesenta}), 
$(c_1(L))^2={2\chi+3\sigma\over4}$, so the index of the
Dirac operator $D:\Gamma(S^{+}\otimes L)\to \Gamma(S^{-}\otimes L)$ is
precisely $\Delta$,  
\be
{\hbox{\rm Index}}~(D)=-{\sigma\over8}+\half (c_1(L))^2={\chi+\sigma\over4}
=\Delta\in
{\IZ}.
\label{cochentatres}
\ee 
 Then,
\be C_2=i^\Delta C_1.
\label{cochentacuatro}
\ee

Finally, we take both contributions and sum over basic
classes. The final form of the generating function (\ref{generador})
turns out to be:
\bea
F(\lambda,\alpha_1,\alpha_2,\dots)
=C_1\left[\ex^{\scriptstyle{\left({\vocalv^2\over 2}  v^2+\langle{\cal
O}\rangle \lambda\right)}}
\sum_x n_x\ex^{\scriptstyle{\langle V\rangle v\cdot x}}+i^\Delta  
\ex^{\scriptstyle{\left(-{\vocalv^2\over 2} v^2-\langle{\cal O}\rangle  
\lambda\right)}}
\sum_x n_x\ex^{\scriptstyle{-i\langle V\rangle v\cdot 
x}}\right].\nonumber\\
\label{cochentacinco}
\eea By comparing to known results by Kronheimer and Mrowka \cite{km} 
the constants $\vocal$ and $\vocalv$ in (\ref{cochentacinco}) are fixed to
be:
\be
 \langle{\cal O}\rangle=2,\qquad
\langle{V}\rangle=1.
\label{cochentaseis}
\ee 
These quantities are universal, \ie,
entirely independent of the manifold $X$. This turns out to be the  case
according to the values (\ref{cochentaseis}), a very important test of 
construction. Different aspects of Seiberg-Witten solution are reflected 
in  (\ref{cochentacinco}). The fact that this formula fits all known 
mathematical results for simply-connected manifolds with $b_2^+>1$ is
rather  satisfactory from the physical point of view.

We are now in a position to write down the final expression for  the
generating function of the Donaldson invariants:
\be
F(\lambda,\alpha_1,\alpha_2,\dots)
=2^{\scriptstyle{1+{1\over4}(7\chi+11\sigma)}}
\left[\ex^{\left({{v^2}\over2}+2\lambda\right)}
\sum_x n_x\ex^{\scriptstyle{v\cdot x}}+i^\Delta  
\ex^{\left(-{v^2\over2}-2\lambda\right)}
\sum_x n_x\ex^{\scriptstyle{-iv\cdot x}}\right].
\label{cochentasiete}
\ee 
The expression above verifies the so-called simple type
condition:
\be
\left({\partial^2\over{\partial \lambda^2}}-4\right)
F(\lambda,\alpha_1,\alpha_2,\dots) =0.
\label{cochentaocho}
\ee All simply-connected four-manifolds with  
$b^{+}_2>1$ for which (\ref{cochentasiete}) is known verify this property.

So far we have discussed two different moduli problems in  four-dimensional
topology, one defined by the ASD instanton  equations and another one
defined by the Seiberg-Witten monopole  equations. There is a natural
generalization of these moduli  problems which involves a non-Abelian
gauge group and also  includes spinor fields. It is the moduli problem
defined by  the  non-Abelian monopole equations, introduced in
Ref. \cite{nabm}  in the context of the  Mathai-Quillen formalism and
as a generalization of Donaldson theory. It has been also considered in
Ref. \cite{ans}\cite{park}, as well as in the  mathematical literature
\cite{oko}\cite{tele}\cite{pt}\cite{oscar}. 

In order to introduce these equations in the case of $G=SU(N)$ and the 
monopole fields in the fundamental representation ${\bf N}$ of $G$,  let
us consider a  Riemannian four-manifold $X$ together  with a principal
$SU(N)$-bundle $P$ and a vector bundle $E$ associated to $P$  through the
fundamental representation.  Suppose for simplicity that the manifold is  
spin, and consider a section $M_{\alpha}^i$ of $S^{+}\otimes  E$. The
non-Abelian monopole equations read in this case: 
\bea && F^{+ij}_{\alpha \beta} + i \Big( {\overline M}^j_{(\alpha}
M^i_{\beta)}- {\delta^{ij} \over N}{\overline M}^k_{(\alpha} M^k_{\beta)}\Big)
 =0, 
\nonumber\\  && (D^{\,\,\dot \alpha \alpha}_{E}M_{\alpha})^i = 0.
\label{namon}
\eea Starting from these equations it is possible to build the associated  
TQFT within the Mathai-Quillen formalism. Not
surprisingly, the  resulting theory is the non-Abelian version of the
topological theory of Abelian  monopoles, that is, a twisted version of
$\cn=2$ super Yang-Mills coupled to one  massless hypermultiplet. The field
content is just the  non-Abelian  version of that of the Abelian monopole
theory. The model can be extended by considering more than one
hypermultiplet ($N_f > 1$), as proposed in \cite{moorewitten}. Let us
briefly describe in this note the case  $N_f = 1$. We will follow the
presentation in \cite{laplata}\cite{nabm}\cite{last}. For a review, see
\cite{tesis}.

From the monopole equations (\ref{namon}) follows that the appropriate 
geometric setting is the following. The field space is ${\cal
A}\times\Gamma(X,S^+\otimes E)$, which is  the space of gauge connections
on
$P$ and positive chirality spinors in the  representation  ${\bf N}$ of
$G$. The vector bundle has as  fiber, ${\cal F}=\Omega^{2,+}(X,\ad
P)\oplus\Gamma(X,S^+\otimes E)$, as dictated by the  quantum numbers of the
monopole equations.  The dimension of the moduli space of non-Abelian
monopoles, ${\cal M}_{\rm NA}$, is provided by a suitable index theorem
\cite{nabm}. It takes the form:
\bea {\rm dim} \,\ {\cal M}_{\rm NA}&=&{\rm dim} \,\ {\cal M}_{\rm  ASD}+2
\,\ {\rm index}\,\ D_E\nonumber\\ &=&(4N-2)c_2(E)-{N^2-1 \over 2}(\chi
+\sigma)-{N
\over 4}\sigma.
\label{cnoventasiete}
\eea  Notice that ${\cal M}_{\rm ASD}\subset{\cal M}_{\rm NA}$. In
addition to this, the usual conditions to have a well-defined moduli
problem (like the reducibility) are essentially the same  as in Donaldson
theory. 

The observables of the theory are the same as in the Donaldson-Witten  theory
since no non-trivial  observables involving matter fields have been found.
The  topological  invariants are then given by correlation functions of
the form  (\ref{clara}). In the perturbative regime, $g\to 0$, one finds the
same pattern as in   ordinary Donaldson-Witten theory. There is a map like
in (\ref{prodi}),
$ H_k(X)\too H^k({\cal M}_{\rm NA}) $, which implies that the vevs of the
theory give a new set of  polynomials in
$H_{k_1}({\cal M}_{\rm NA})\times H_{k_2}({\cal M}_{\rm NA})\times\ldots\times
H_{k_p}({\cal M}_{\rm NA})$. As in  the case of the  Donaldson-Witten
theory, the perturbative approach does not  provide any further insight into the
precise form of these topological  invariants. Fortunately, it is again possible
to apply the results of Seiberg and  Witten  on $\cn=2$ supersymmetric theories
to analyze the model at hand in the non-perturbative regime.

To carry out the analysis at strong coupling one can follow the same
strategy as in the case of Donaldson-Witten theory. The physical theory
underlying the theory of non-Abelian monopoles  is an 
$\cn=2$ supersymmetric Yang-Mills theory coupled to one massless
hypermultiplet in the fundamental representation of the gauge group, which
we take to be  
$SU(2)$.  This theory is asymptotically free. Hence, it is weakly coupled 
($g\to 0$) in the ultraviolet, and strongly coupled ($g\to
\infty$) in  the infrared. The infrared behavior of this theory has been 
also determined by Seiberg and Witten \cite{sw}. As in the previous case
the quantum moduli space of vacua is a one-dimensional complex
K\"ahler manifold (the $u$-plane) and for any $u$ there is an unbroken
$U(1)$ gauge symmetry (Coulomb  phase).
At a generic point on the $u$-plane the only light degree of 
freedom is the $U(1)$ gauge field (together with its $\cn=2$ superpartners). 
However, in this case there are three singularities at finite values of $u$.
These values are: $u_1=-1$, $u_2=\ex^{-i{\pi\over3}}$ and
$u_3=\ex^{i{\pi\over3}}$. Near each of these singularities  a magnetic
monopole or dyon  becomes massless and weakly coupled to a dual
$U(1)$ gauge field.

For $X$ such that $b^+_2>1$ (there are no Abelian instantons) the only 
contributions come from the three singularities. Following the same
arguments  as in the previous case, and assuming that the manifold $X$ is
of simple type, one finds that the contribution from the singularity $u_1$
takes the form:
\bea
\ll\langle \ex^{\scriptstyle{\left(\sum_a \alpha_a  I(\Sigma_a)+\lambda{\cal
O}\right)}}\rr\rangle_{u_1}&=&
C_1\sum_{x} n_x \exp\Bigg(\lambda\vocal u_1  + {i \vocalv \over 2}  v \cdot x
\ll({du\over d a}\rr)_{u_1} +\vocalv^2 v^2 \tilde T(u_1) \Bigg)
\nonumber \\
&=& C_1\sum_{x} n_x \exp\Big(-\lambda\vocal   + i \vocalv \sqrt{2}  v \cdot x
 -{2\over 3} \vocalv^2 v^2 \Big),
\label{cnoventanueve}
\eea 
where,
\be
\tilde T(u) = -{1\over 24} \ll(\ll({du\over da}\rr)^2 - 8 u\rr),
\label{lattilde}
\ee
and the quantities $C$, $\vocal$ and $\vocalv$ are constants. In
(\ref{cnoventanueve}) we have used the fact that
\be
\tilde T(u_j) = {2\over 3} u_j; \,\,\,\,\,\,\,\,\,\,\,\,
\ll( {du\over d a_D} \rr)_{u_j}^2 = -8u_j.
\label{pluf}
\ee
The contributions from the other singular points are obtained using
the  broken
$U(1)_{\cal R}$ symmetry which in this case is ${\IZ}_6$. This symmetry
generates a ${\IZ}_3$ symmetry on the $u$-plane which acting on the
observables takes the form:
\bea &&I(\Sigma_a)\too  
\ex^{{-2i\pi\over3}}I(\Sigma_a),\nonumber\\  &&{\cal O}\too
\ex^{{2i\pi\over3}}{\cal O},
\label{dtres}
\eea 
under the action of the  generator of ${\IZ}_3$. The contribution from
each singularity possess a relative global factor which is obtained from
the form of the gravitational anomalies \cite{laplata}\cite{last}. The final 
form
of the generating function (\ref{generador}) for manifolds of simple is
\cite{last}\cite{moorewitten}:
\begin{eqnarray}  F(\lambda,\alpha_1,\alpha_2,\dots) &=&C
\Bigg( {\rm exp} \Big(- {2\over 3} \vocalv^2 v^2  - \lambda \vocal\Big)
\sum_{x} n_x {\rm exp}(i\sqrt{2} \vocalv v \cdot x)  \nonumber \\ &+&{\rm
e}^{-i{\pi 
\over 6}\sigma} {\rm exp}\Big( {\rm e}^{-{\pi  \over 3}i} ({2\over 3}
\vocalv^2 v^2  + \lambda \vocal)
\Big) \sum_{x} n_x {\rm exp}({\rm e}^{{i\pi  \over 3}i} \sqrt{2} \vocalv
v \cdot x) \nonumber \\ &+&{\rm e}^{-i{\pi  \over 3}\sigma} {\rm exp}\Big(
{\rm e}^{{\pi  \over 3}i} ({2\over 3}
\vocalv^2 v^2  + \lambda \vocal)
\Big) \sum_{x} n_x {\rm exp}({\rm e}^{-{i\pi  \over 3}i} \sqrt{2} \vocalv
v \cdot x) \Bigg), \nonumber\\
\label{fernanda}
\end{eqnarray} where unknown constants appear as in the pure
Donaldson-Witten case.  The generating function (\ref{fernanda}) verifies
a generalized form of the simple type condition (\ref{cochentaocho}):
\be 
\left({\partial^3\over{\partial\lambda^3}}-\vocal^3\right)
\ll\langle {\rm exp} (\sum_{a}\alpha_{a}I(\Sigma_{a})+\lambda {\cal  O})
\rr\rangle =0.
\label{docho}
\ee
Unfortunately, the left-hand side of (\ref{fernanda}) is not known for any
manifold $X$. Thus we can not fix the unknown constants as we did in the
case of Donaldson theory (but see \cite{geog}, where a general recipe to partly
determine these constants is proposed.)

Generalized Donaldson-Witten theory for $N_f>1$ has been considered in
\cite{moorewitten}\cite{geog}\cite{kanoyang}. For $N_f<4$ results similar to
(\ref{fernanda}) are obtained for the case of manifolds of simple type. We will
not review these cases in this paper. Instead we will turn our attention to the
case  of some of the twisted theories which emerge from $\cn=4$ supersymmetric
theories.

 Unlike the 
$\cn=2$ supersymmetric case, $\cn=4$ supersymmetric Yang-Mills theory is
unique once the gauge group
$G$ is fixed. The  microscopic theory contains a gauge or gluon field, 
four chiral spinors (the gluinos) and six real scalars. All the above fields 
are massless and take values in the adjoint representation of the gauge group. 
As in the $\cn=2$ case, the ${\cal R}$-symmetry group of the $\cn=4$ algebra 
can be twisted to obtain a topological model. But since the 
${\cal R}$-symmetry is now $SU(4)$, this topological twist can be performed in 
three  inequivalent ways, so one ends up with three different TQFTs
\cite{vafawitten}\cite{yamron}\cite{ene4}. The twisted theories are  topological
in the sense that the partition  function as well as a selected set of
correlation functions are independent  of the metric which defines the
background geometry. In the short distance  regime, computations in the twisted
theory are given exactly by a saddle-point  calculation around a certain bosonic
background or moduli space, and in fact  the correlation functions can be
reinterpreted  as describing intersection theory on this moduli space. This
correspondence can  be made more precise through the Mathai-Quillen construction
\cite{ene4}.  Unfortunately, it is not possible to 
perform explicit computations from this viewpoint: the moduli spaces one ends 
up with are generically non-compact, and no precise recipe is known to 
properly compactify them. 

As explained above, a complementary approach which sheds more light on the 
structure of the 
twisted theories and allows explicit computations involves the long-distance 
regime (or strong coupling regime in the asymptotically free theories), where 
one expects that a good description should be provided by the 
degrees of freedom of the vacuum states of the physical theory on $\IR^4$. 
We have seen above how this program works for the $\cn=2$ theories, and it 
would be interesting to see whether similar constructions work for the $\cn=4$ 
theories. 

While for the 
TQFTs related to asymptotically free ${\cal N}=2$ theories the 
interest lies in their ability to define topological invariants for 
four-manifolds, for the twisted ${\cal N}=4$ theories the topological character 
is used as a tool for performing explicit computations which might shed light on 
the structure of the physical ${\cal N}=4$ theory. This theory is finite and 
conformally invariant, and is conjectured to have a symmetry exchanging strong 
and weak coupling and exchanging electric and magnetic fields, which extends to 
a full $SL(2,\IZ)$ symmetry acting on the microscopic complexified coupling 
$\tau_0$ \cite{monoli}. It is natural to expect that this property should be 
shared by the 
twisted theories on arbitrary four-manifolds. This was checked by Vafa and 
Witten for one of the twisted theories and for gauge group $SU(2)$ 
\cite{vafawitten}, and it was 
clearly mostly interesting to extend their computation to higher rank groups 
and to the other twisted theories. 

In \cite{htwist} the $u$-plane 
approach was applied to the twisted mass-deformed ${\cal N}=4$ SYM theory with 
gauge group $SU(2)$. This theory is obtained by twisting the ${\cal N}=4$ SYM 
theory with bare masses for two of the chiral multiplets. It is a non-Abelian 
monopole theory as the one described above but with the monopole multiplets 
taking values in the adjoint representation of the gauge group \cite{ene4}.
The physical theory preserves ${\cal N}=2$ supersymmetry, and its low-energy 
effective description for gauge group $SU(2)$ was given by Seiberg and Witten 
\cite{sw}, and later extended to $SU(N)$ in \cite{donagi}. 

The ghost-number symmetry of the twisted theory for gauge group $SU(2)$ 
has an anomaly $-3(2\chi+3\sigma)/4$ on gravitational backgrounds. Topological 
invariants are thus obtained by considering the vacuum expectation value of 
products of observables with ghost-numbers adding up to $-3(2\chi+3\sigma)/4$. 
The relevant observables for this theory and gauge group $SU(2)$ or $SO(3)$ are 
precisely the same as in the Donaldson-Witten theory (\ref{raquela}) and 
(\ref{raquel}). 
In addition to this, since all the fields in the theory take values in the 
adjoint representation of the gauge group, it is possible to enrich the 
theory by including non-Abelian electric and magnetic 't Hooft fluxes 
\cite{gthooft} which should behave under $SL(2,\IZ)$ duality in a well-defined 
fashion \cite{vafawitten}\cite{gthooft}.

The generating function for these correlation functions is given \cite{htwist}
as an integration over the moduli space of vacua (the $u$-plane) of the 
physical theory. At a generic vacuum, the only contribution comes from a 
twisted $\cn=2$ Abelian 
vector multiplet. The effect of the massive modes is contained in appropriate 
measure factors, which also incorporate the coupling 
to gravity, and in contact terms among the observables 
\cite{moorewitten}\cite{lns}\cite{mmone}.  

The total contribution to the generating function thus consists of an 
integration 
over the moduli space with the singularities removed -- which is 
non-vanishing for $b^{+}_2(X)=1$ \cite{moorewitten} only -- plus a discrete sum 
over the contributions of the twisted effective theories at each of the 
three singularities of the low-energy effective description \cite{sw}. 
The effective theory at a given singularity contains, 
together with the appropriate dual photon multiplet, one charged 
hypermultiplet, which corresponds to the state becoming massless at 
the singularity. The complete effective action for these 
massless states contains as well certain measure factors and 
contact terms among the observables, which reproduce the 
effect of the massive states which have 
been integrated out. How to determine these a priori 
unknown functions was explained in \cite{moorewitten}. 
The idea is as follows. At those points on the $u$-plane where the 
(imaginary part of the) effective coupling 
diverges, the integral is discontinuous at antiself-dual Abelian 
gauge configurations. This is commonly referred to as 
``wall crossing". Wall crossing can take place at the 
singularities of the moduli space -- the appropriate local effective 
coupling $\tau$ diverges there -- and, in the case of the 
asymptotically free theories, at the point at infinity -- 
the effective electric coupling diverges owing to asymptotic 
freedom.

On the other hand, the final expression for the invariants can 
exhibit a wall-crossing behavior at most at $u\to\infty$, so the 
contribution to  wall crossing from the integral at the 
singularities at finite values of $u$ 
must cancel against the contributions coming from the effective theories 
there, which also display wall-crossing discontinuities.  
Imposing this cancellation fixes almost completely the 
unknown functions in the contributions to the topological correlation 
functions from the singularities. The final result for the contributions 
from the singularities  (which give the complete  answer for the correlation 
functions when $b^{+}_{2}>1$) is written explicitly and completely in terms 
of the periods and the discriminant of the 
Seiberg-Witten solution for the physical theory. For simply-connected spin 
four-manifolds of simple type the generating function is given by:
\begin{eqnarray}
\left\langle\ex^{p{\cal O}+I(S)}\right\rangle_{v}= 2^{{\nu\over2}+{2\chi+
3\sigma\over8}} m^{-(3\nu+\sigma/4)} (\eta(\tau_0))^{-12\nu}& &
\hbox{\hskip-15pt}\Bigg\{
\nonumber\\
(\kappa_1)^{\nu} \left ({{da}\over{du}}\right)^
{-(\nu+{\sigma\over4})}_1\ex^{2pu_1 + S^2 T_1} & &\hbox{\hskip-15pt}\sum_{x} 
\delta_{\left[{x\over2}\right],v} \,n_x \,\ex^{{i\over2} \left (du
/da\right)_1 x\cdot S}\nonumber\\ +
2^{-{b_2\over2}}(-1)^{\sigma/8}(\kappa_2)^{\nu} \left 
({{da}\over{du}}\right)^
{-(\nu+{\sigma\over4})}_2\,\ex^{2pu_2 + S^2 T_2} & &\hbox{\hskip-15pt}\sum_{x} 
(-1)^{v \cdot {x\over2}}\,n_x\, \ex^{{i\over2} \left (du/
da\right)_2 x\cdot S}\nonumber\\ +
2^{-{b_2\over2}} i^{-v^2}(\kappa_3)^{\nu} \left 
({{da}\over{du}}\right)^
{-(\nu+{\sigma\over4})}_3\ex^{2pu_3 + S^2 T_3}& &\hbox{\hskip-15pt}\sum_{x} 
(-1)^{v \cdot {x\over2}}\,n_x\, \ex^{{i\over2} \left (du/
da\right)_3 x\cdot S}\Bigg\},
\label{formulauno}
\end{eqnarray}
where $x$ is a Seiberg-Witten basic class, $\nu=(\chi+\sigma)/4$, $v\in
H^2(X,\IZ_2)$  is a 't Hooft flux, $\eta(\tau_0)$ is the Dedekind function, 
$\kappa_i=(du/dq)_{u=u_i}$ -- with $q=\exp(2\pi i \tau)$ -- and the contact terms
$T_i$ have the form 
\be
T_i= -{1 \over 12} \left({du\over da}\right)_{u=u_i}^{\,\,\, 2} +
E_2 (\tau_0) {u_i \over 6} + {m^2 \over 72} E_4 (\tau_0) 
\label{ender}
\ee
being $E_2$ and $E_4$ are 
the Eisenstein series of weights $2$ and $4$ respectively\footnote{Notice that
we have changed the notation used in (\ref{generador}). The parameter $\lambda$
has been replaced by $p$ and the formal sum $v=\sum_a \alpha_a \Sigma_a$
by
$S$. In the rest of the paper $v$ will denote a 't Hooft flux.}. Evaluating
the quantities in  (\ref{formulauno}) gives the final result as a function of
the physical  parameters $\tau_0$ and
$m$, and of topological data of $X$  as $\chi$, $\sigma$ and the basic classes
$x$. 

The formula has nice properties under the modular group. For the partition 
function $Z_v$,
\begin{eqnarray}
Z_v(\tau_0+1) &=& (-1)^{\sigma/8}i^{-v^2} Z_v(\tau_0), \nonumber\\
Z_v(-1/\tau_0) &=& 2^{-b_2/2}(-1)^{\sigma/8}
\left({\tau_0\over i}\right)^{-\chi/2}
\sum_w (-1)^{w\cdot v} Z_w(\tau_0). 
\label{dualityi}
\end{eqnarray}

Also, with  $Z_{SU(2)}= 2^{-1} Z_{v=0}$ and $Z_{SO(3)}=
\sum_v Z_v$, 
\begin{eqnarray}
Z_{SU(2)}(\tau_0+1)&=&(-1)^{\sigma/8}Z_{SU(2)}(\tau_0),\nonumber\\
Z_{SO(3)}(\tau_0+2)&=& Z_{SO(3)}(\tau_0),\nonumber\\
Z_{SU(2)}(-1/\tau_0)&=&(-1)^{\sigma/8}2^{-\chi/2}\tau_{0}{}^{-\chi/2}
Z_{SO(3)}(\tau_0).
\label{tdual}
\end{eqnarray}
Notice that the last of these 
three equations corresponds precisely to the strong-weak coupling 
duality transformation conjectured by Montonen and Olive \cite{monoli}.

As for the correlation functions, one finds the following behavior under 
the inversion of the coupling
\begin{eqnarray}
\left\langle{1\over8\pi^2}\tr\,\phi^2\right\rangle
^{SU(2)}_{\tau_0}&=&\left\langle {\cal O}
\right\rangle^{SU(2)}_{\tau_0}={1\over\tau_0{}^2}\left\langle 
{\cal O}\right\rangle^{SO(3)}_{-1/\tau_0},\nonumber\\ \nonumber\\
\left\langle{1\over8\pi^2}\int_S\tr\,\left(2\phi F + \psi\wedge\psi\right)
\right\rangle^{SU(2)}_{\tau_0}&=&
\left\langle I(S)\right\rangle^{SU(2)}_{\tau_0}= 
{1\over\tau_0{}^2}\left\langle I(S)\right\rangle^{SO(3)}_{-1/\tau_0},
\nonumber\\ \nonumber\\
\left\langle I(S)I(S)\right\rangle^{SU(2)}_{\tau_0}\,\,\,= 
\left({\tau_0\over i}\right)^{-4}& & \!\!\!\!\!\!\!\!\!\!\!\!\!
\left\langle I(S)I(S)\right
\rangle^{SO(3)}_{-1/\tau_0}+{i\over2\pi}{1\over\tau_0{}^3}
\left\langle {\cal O}\right\rangle^{SO(3)}_{-1/\tau_0}\sharp 
(S\cap S).
\label{bizarre}
\end{eqnarray}
Therefore we see that, as expected, the partition function of the twisted 
theory transforms as a modular form, while the topological correlation 
functions  turn out to transform covariantly under $SL(2,\IZ)$, following a 
pattern which can be reproduced with a far more simple topological Abelian 
model \cite{htwist}.

The second example we will consider is the Vafa-Witten theory 
\cite{vafawitten}, which corresponds to another non-equivalent twist of
the 
$\cn=4$ theory. The twisted theory does not contain spinors, so it is 
well-defined on any compact, oriented four-manifold. The ghost-number
symmetry  of this theory is anomaly-free, and therefore the only
non-trivial topological  observable is the partition function itself. As
in the above example, it is  possible to consider non-trivial gauge
configurations in $G/Center(G))$ and  compute the partition function for a
fixed value of the 't Hooft flux $v\in  H^{2}(X,\pi_1(G))$. In this case,
however, the Seiberg-Witten   approach is not available, but, as
conjectured by Vafa and Witten, one can  nevertheless compute in terms of
the vacuum degrees of freedom of the 
${\cal N}=1$ theory which results from giving bare masses to all the three 
chiral multiplets of the ${\cal N}=4$ theory\footnote{A similar approach
was  introduced by Witten in \cite{wijmp} to obtain the first explicit
results for  the Donaldson-Witten theory just before the far more powerful
Seiberg-Witten  approach were available.}. 

As explained in detail in
\cite{vafawitten}\cite{masas}\cite{coreatres}, the twisted massive theory is
topological on K\"ahler four-manifolds  with $h^{2,0}\not=0$, and the
partition function is actually invariant under  the perturbation. In the
long-distance limit, the partition function is   given as a finite sum
over the contributions of the discrete massive vacua of  the resulting
$\cn=1$ theory. In the case at hand, it turns that for $G=SU(N)$,  the
number of such vacua is given by the sum of the positive divisors of $N$ 
\cite{donagi}. The contribution of each vacuum is universal (because of the mass
gap), and can be fixed by comparing to known mathematical results
\cite{vafawitten}. However, this is not the end of the story. In the
twisted theory  the chiral superfields of the $\cn=4$ theory are no longer
scalars, so the  mass terms can not be invariant under the holonomy group
of the manifold  unless one of the mass parameters be a holomorphic
two-form $\omega$.\footnote {Incidentally, this is the origin of the
constraint $h^{(2,0)}\not=0$ mentioned above.} This spatially dependent
mass term vanishes where $\omega$ does, and  we will assume as in
\cite{vafawitten}\cite{wijmp} that $\omega$  vanishes with
multiplicity one on a union of disjoint,  smooth complex curves $C_i$,
$i=1,\ldots n$ of genus $g_i$ which represent the  canonical divisor $K$
of $X$. The vanishing of $\omega$ introduces corrections  involving $K$ whose
precise form is not known a priori. In the $G=SU(2)$ case,  each of the
$\cn=1$ vacua bifurcates along each of the components $C_i$ of the 
canonical divisor into two strongly coupled massive vacua. This vacuum 
degeneracy is believed to stem \cite{vafawitten}\cite{wijmp} from the
spontaneous  breaking of a $\IZ_2$ chiral symmetry which is unbroken in
bulk. 

The structure of the corrections for $G=SU(N)$ -- see (\ref{partidos})
below --  suggests that the mechanism at work in this case is not chiral
symmetry  breaking. Indeed, near any of the $C_i$ there is an $N$-fold
bifurcation of the  vacuum. A plausible explanation for this degeneracy
could be found in the  spontaneous breaking of the center of the gauge
group (which for $G=SU(N)$ is  precisely $\IZ_N$). For further details 
and speculations, we refer the reader to \cite{sun}. In any case, the
formula  for $SU(N)$ can be computed (at least when $N$ is prime) along
the lines  explained in \cite{vafawitten} and assuming that the resulting
partition  function satisfies a set of non-trivial constraints which are 
described below.   


Then, for a given 
't Hooft flux $v\in H^{2}(X,\IZ_{N})$, the partition function for gauge group 
$SU(N)$ (with prime $N$ ) is \cite{sun}:
\begin{eqnarray}
Z_v &=& \left (\sum_{\vec\varepsilon}
\delta_{v,w_N(\vec\varepsilon)}
\prod_{i=1}^{n}\prod_{\lambda=0}^{N-1} 
\left({\chi_{\lambda}\over 
\eta}\right)^{(1-g_i)\delta_{\varepsilon_i,\lambda}}\right)
\left ({1\over N^2}
G(q^N)\right)^{\nu/2} \nonumber\\
&+& N^{1-b_1} \sum_{m=0}^{N-1}\left[\prod_{i=1}^{n}
\left( \sum_{\lambda=0}^{N-1} 
\left({\chi_{m,\lambda}\over \eta}\right)^{1-g_i} 
\ex^{{2i\pi\over N}\lambda v\cdot[C_i]_N}\right)\right]
\ex^{i\pi{N-1\over N}m v^2}\left( {1\over N^2}
G(\alpha^m q^{1/N})\right)^{\nu/2}, \nonumber\\
\label{partidos}
\end{eqnarray}
where $\alpha={\hbox{\rm exp}}(2\pi i/N)$, $G(q)=\eta(q)^{24}$ (with $\eta(q)$ 
the Dedekind function), $\chi_\lambda$ are the $SU(N)$ characters at level $1$ 
\cite{izzy} 
and $\chi_{m,\lambda}$ are certain linear combinations thereof. 
$[C_i]_N$ is the reduction modulo $N$ of the Poincar\'e dual of $C_i$, 
and 
\be
w_N(\vec\varepsilon)=\sum_{i=1}^n \varepsilon_i [C_i]_N,
\label{flux}
\ee
where $\varepsilon_i=0,1,\ldots N-1$ are chosen independently.

The formula (\ref{partidos}) does not apply directly to the $N=2$ case. For 
$N=2$ there are some extra relative phases $t_i$ -- see equations (5.45) and
(5.46)  in \cite{vafawitten} -- which are absent for $N>2$ and prime. Modulo
these extra  phases, (\ref{partidos}) is a direct generalization of Vafa and
Witten's  formula. It reduces on $K3$ to the formula of Minahan, 
Nemeschansky, Vafa and Warner \cite{estrings} and generalizes their result to 
non-zero 't Hooft flux. 
In addition to this, the formula has the expected properties under the modular 
group \cite{vafawitten}
\begin{eqnarray}
Z_v(\tau_0+1) &=&\ex^{{i\pi\over12}N(2\chi+3\sigma)}\ex^{-i\pi{N-1\over N}v^2}
Z_v(\tau_0), \nonumber\\
Z_v(-1/\tau_0) &=& N^{-b_2/2}\left({\tau_0\over i}\right)^{-\chi/2}
\sum_u \ex^{{2i\pi u\cdot v\over N}} Z_u(\tau_0), 
\label{duality}
\end{eqnarray}
and also, with  $Z_{SU(N)}= N^{b_1-1} Z_{0}$ and $Z_{SU(N)/\IZ_N}=
\sum_v Z_v$, 
\be
Z_{SU(N)}(-1/\tau_0)=  
N^{-\chi/2}\left({\tau_0\over i}\right)^{-\chi/2}Z_{SU(N)/\IZ_N}(\tau_0),
\label{montolive}
\ee
which is, up to some correction factors which vanish in flat space, 
the original Montonen-Olive conjecture. 

There is a further property to be checked which concerns the behavior of 
(\ref{partidos}) under blow-ups. This property was heavily used in 
\cite{vafawitten} and demanding it in the present case was essential in 
deriving the above formula. Blowing up a point on a K\"ahler manifold $X$ 
replaces it 
with a new K\"ahler manifold $\hat X$ whose second cohomology lattice is 
$H^{2}({\hat X},\IZ)= H^{2}({X},\IZ)\oplus I^{-}$, where $I^{-}$ is the 
one-dimensional lattice spanned by the Poincar\'e dual of the exceptional
divisor 
$B$ created by the blow-up. Any allowed $\IZ_N$ flux $\hat v$ on $\hat X$ 
is of the form $\hat v=v\oplus r$, where $v$ is a 
flux in $X$ and $r=\lambda B$, $\lambda=0,1,\ldots N-1$. The main result 
concerning (\ref{partidos}) is that under blowing up a point on a K\"ahler 
four-manifold with canonical divisor as above, the partition functions for fixed 
't Hooft fluxes have a factorization as 
\be
Z_{\hat X,\hat v}(\tau_{0})= Z_{X,v}(\tau_{0})\,\,{\chi_{\lambda}(\tau_{0})
\over \eta(\tau_{0})}.
\label{factoriz}
\ee
Precisely the same behavior under blow-ups of the partition function 
(\ref{partidos}) has been proved by Yoshioka \cite{yoshioka} for 
the generating function of Euler characteristics of instanton moduli space 
on K\"ahler manifolds. This should not come out as a surprise since it is known 
that, on certain four-manifolds, the partition function of 
Vafa-Witten theory computes Euler characteristics of 
instanton moduli spaces \cite{vafawitten}\cite{estrings}. Therefore,
(\ref{partidos}) can be seen as a prediction for the Euler
numbers of instanton moduli spaces on those four-manifolds. 

In the light of the AdS/CFT correspondence
\cite{malda}\cite{klebanov}\cite{wittenads}, it would be mostly interesting to 
investigate what the large $N$ limit of (\ref{partidos}) corresponds to on the 
gravity side, and to extend the computation to all $N$. We expect to address 
these topics in the near future.

\vskip2cm
\begin{center} {\bf Acknowledgments}
\end{center}

\vspace{4 mm}

We would like to thank M. Mari\~no for helpful
discussions. J.M.F.L. would like to thank the organizers of
the workshop on ``New Developments in Algebraic Topology" for their kind
invitation and their hospitality. This work was supported in part by DGICYT
under grant PB96-0960, and by the EU Commission under TMR grant
FMAX-CT96-0012.


\vfill
\newpage



\begin{thebibliography}{99}


\def\np{Nucl. Phys.}
\def\pl{Phys. Lett.} 
\def\pre{Phys. Rep.} 
\def\prl{Phys. Rev. Lett.}
\def\pr{Phys. Rev.} 
\def\ap{Ann. Phys.} 
\def\cmp{Comm. Math. Phys.}
\def\ijmp{Int. J. Mod. Phys.} 
\def\mpl{Mod. Phys. Lett.} \def\lmp{Lett. Math. Phys.} 
\def\bams{Bull. AMS} \def\am{Ann. of Math.} 
\def\jpsc{J. Phys. Soc. Jap.} \def\topo{Topology} 
\def\ijm{Int. J. Math.}
\def\knot{Journal of Knot Theory and Its Ramifications} 
\def\jmp{J. Math. Phys.} 
\def\jgp{J. Geom. Phys.} 
\def\jdg{J. Diff. Geom.}
\def\plms{Proc. London Math. Soc.}
\def\mrl{Math. Res. Lett.}
\def\inma{Invent. Math.}
\def\tam{Trans. Am. Math. Soc.}

\bibitem{tqft}
E. Witten,
``Topological quantum field theory,"
{\sl Commun. Math. Phys.} {\bf 117} (1988)
353.

\bibitem{thompson}
D. Birmingham, M. Blau, M. Rakowski and G.
Thompson, ``Topological field theories," {\sl Phys. Rep.} 
{\bf 209} (1991) 129.

\bibitem{moore} S. Cordes, G. Moore and S. Rangoolam, 
``Lectures on  2D Yang-Mills theory, equivariant cohomology and
topological field theory," in {\it Fluctuating geometries in statistical 
mechanics and field theory}, Les Houches Session LXII, F. David, P. Ginsparg 
and J. Zinn-Justin, eds. (Elsevier, 1996) p. 505; hep-th/9411210.

\bibitem{laplata} J.M.F. Labastida and Carlos Lozano, ``Lectures on
topological  quantum field theory," in Proceedings of the CERN-Santiago de 
Compostela-La Plata Meeting on ``Trends in Theoretical Physics", 
H. Falomir, R. Gamboa, 
F. Schaposnik, eds. (American Institute of Physics, New York, 1998); 
hep-th/9709192.

\bibitem{sw} 
N. Seiberg and E. Witten,
``Electric-magnetic duality, monopole condensation, and confinement in
${\cal N}=2$ supersymmetric Yang-Mills
Theory,"
hep-th/9407087, {\sl Nucl. Phys.} {\bf B426} (1994) 19, Erratum, {\sl 
ibid} {\bf B430} (1994) 485; 
``Monopoles, Duality and chiral symmetry breaking in ${\cal N}=2$ 
supersymmetric QCD,"
hep-th/9408099; {\sl Nucl. Phys.} {\bf B431} (1994) 484.

\bibitem{donald}S.K. Donaldson, {\sl\topo} {\bf 29} (1990) 257.

\bibitem{malda}
J. Maldacena, ``The large $N$ limit of superconformal field theories 
and supergravity," {\sl Adv. Theor. Math. Phys.} {\bf 2} (1998) 231; 
hep-th/9711200.

\bibitem{klebanov}
S.S. Gubser, I. Klebanov and A.M. Polyakov, ``Gauge theory correlators from
non-critical string theory," {\sl Phys. Lett.} {\bf B428} (1998) 105;
hep-th/9802109.

\bibitem{wittenads}
E. Witten, ``Anti-de Sitter space and holography," 
{\sl Adv. Theor. Math. Phys.} {\bf 2} (1998) 253; hep-th/9802150.

\bibitem{hull}
C.M. Hull, ``Timelike T-duality, de Sitter space, large $N$ gauge 
theories and topological field theory," 
{\sl J. High Energy Phys.} {\bf 9807} (1998) 021; hep-th/9806146.

\bibitem{gopaku}
R. Gopakumar and C. Vafa, ``Topological gravity as large $N$ 
topological gauge theory,"  
{\sl Adv. Theor. Math. Phys.} {\bf 2} (1998) 413, hep-th/9802016; 
``On the gauge theory/geometry 
correspondence," hep-th/9811131.

\bibitem{jeffrey}
M.F. Atiyah and L. Jeffrey, ``Topological Lagrangians 
and cohomology," {\sl\jgp} {\bf 7} (1990) 119.

\bibitem{mathai}
V. Mathai and D. Quillen, {\sl Topology} {\bf 25} (1986) 85.  



\bibitem{abm}
E. Witten, ``Monopoles and
four-manifolds,"  hep-th/9411102; {\sl Math. Res. Letters} {\bf 1} 
(1994) 769.

\bibitem{rocek} A. Karlhede and M. Ro\v cek, {\sl\pl} {\bf B212} (1988) 51.

\bibitem{top} 
M. Alvarez and J.M.F. Labastida, {\sl\pl} {\bf B315} (1993) 251;
{\sl\np} {\bf B437} (1995) 356.

\bibitem{ans} 
D. Anselmi and P. Fr\`e, {\sl\np} {\bf B392} (1993) 401;
{\sl\np} {\bf B404} (1993) 288; {\sl\np}
{\bf B416} (1994) 25; {\sl\pl} {\bf B347} (1995) 247.

\bibitem{abmono} J.M.F. Labastida and M. Mari\~no, ``A topological Lagrangian 
for monopoles on four-manifolds," {\sl\pl} {\bf B351} (1995) 146; hep-th/9503105.

\bibitem{nabm}
J.M.F. Labastida and  M. Mari\~no, ``Non-Abelian 
monopoles on four-manifolds," {\sl Nucl. Phys.} {\bf B448} (1995) 373; 
hep-th/9504010.

\bibitem{last}
J.M.F. Labastida and M.
Mari\~no, ``Polynomial invariants for $SU(2)$ monopoles," {\sl Nucl. Phys.} 
{\bf B456} (1995) 633; hep-th/9507140.

\bibitem{moorewitten}
G. Moore and E. Witten, ``Integration over
the $u$-plane in Donaldson theory," {\sl Adv. Theor. Math. Phys.} 
{\bf 1} (1998) 298; hep-th/9709193.

\bibitem{lns}
A. Losev, N. Nekrasov, and S. Shatashvili, ``Issues in
topological gauge theory," hep-th/9711108; {\sl Nucl. Phys.} 
{\bf B534} (1998) 549.``Testing Seiberg-Witten solution,"
hep-th/9801061.

\bibitem{mmone}
M. Mari\~no and G. Moore, ``Integrating over the Coulomb branch in
${\cal N}=2$ gauge theory," hep-th/9712062; {\sl Nucl. Phys. B (Proc. Suppl.)} 
{\bf 68} (1998) 336; ``Donaldson invariants for nonsimply connected manifolds," 
hep-th/9804104.

\bibitem{mooremari}
M. Mari\~no and G. Moore, ``The Donaldson-Witten function for gauge
groups
of rank larger than one," {\sl Commun. Math. Phys.} {\bf 199} (1998) 
25; hep-th/9802185.

\bibitem{vafawitten}
C. Vafa and E. Witten,
``A strong coupling test of $S$-duality,"
hep-th/9408074; {\sl Nucl. Phys.} {\bf B431} (1994) 3.


\bibitem{baryon} 
J.M F. Labastida and M. Mari\~no, ``Twisted baryon 
number in ${\cal N}=2$ supersymmetric QCD," {\sl\pl} {\bf B400} (1997) 
323; hep-th/9702054.

\bibitem{wijmp}
E. Witten, ``Supersymmetric Yang-Mills theory
on a four-manifold,"  hep-th/9403193;
{\sl J. Math. Phys.} {\bf 35} (1994) 5101.


\bibitem{llatas} J.M.F. Labastida and P.M. Llatas, {\sl\np} {\bf B379}
(1992) 220.

\bibitem{gian}  R. Gianvittorio ,  I. Martin and  A. Restuccia, {\sl Lett.
Math. Phys.} {\bf 39} (1997) 51.

\bibitem{km} P.B. Kronheimer and T.S. Mrowka, {\sl \bams} {\bf 30} (1994) 215.

\bibitem{park}
S. Hyun, J. Park and J.-S. Park, ``Topological QCD," {\sl Nucl. 
Phys.} {\bf B453} (1995) 199; hep-th/9503020.

\bibitem{oko}
C. Okonek and A. Teleman, {\it Int. J. Math.} {\bf 6} (1995) 893; 
{\sl\cmp} {\bf 180} (1996) 363.

\bibitem{tele}
A. Teleman, ``Non-Abelian Seiberg-Witten theory and projectively 
stable pairs," alg-geom/9609020. 

\bibitem{pt}
V. Pidstrigach and A. Tyurin, ``Localisation of the Donaldson invariants 
along Seiberg-Witten classes," dg-ga/9507004.

\bibitem{oscar}
S. Bradlow and O. Garc\'\i a-Prada, ``Non-Abelian monopoles and vortices," 
alg-geom/9602010.

\bibitem{tesis}
M. Mari\~no, ``The geometry of supersymmetric gauge theories 
in four dimensions," Ph.D. Thesis, Universidade de Santiago de Compostela, 
October 1996; hep-th/9701128.

\bibitem{geog} 
M. Mari\~no, G. Moore and G. Peradze, ``Superconformal 
invariance and the geography of four-manifolds," hep-th/9812055; 
``Four-manifold geography 
and superconformal symmetry," math.DG/9812042.

\bibitem{kanoyang} H. Kanno and S.-K. Yang,
``Donaldson-Witten functions of massless $N$=2 supersymmetric QCD," 
hep-th/9806015; {\sl Nucl.Phys.} {\bf B535} (1998) 512.

\bibitem{yamron}
J.P. Yamron, ``Topological actions in twisted supersymmetric
theories," {\sl Phys. Lett.} {\bf B213} (1988) 325. 

\bibitem{ene4}
J.M.F. Labastida and Carlos Lozano, ``Mathai-Quillen formulation 
of twisted $\cn=4$ supersymmetric gauge theories in four 
dimensions," {\sl Nucl. Phys.} {\bf B502} (1997) 741; hep-th/9702106.

\bibitem{monoli}
C. Montonen and D. Olive, ``Magnetic monopoles as gauge 
particles?" {\sl Phys. Lett.} {\bf B72} (1977) 117.

\bibitem{htwist}
J.M.F. Labastida and Carlos Lozano, ``Duality in twisted ${\cal N}=
4$ supersymmetric gauge theories in four dimensions," {\sl Nucl. Phys.} 
{\bf B537} (1999) 203; hep-th/9806032.

\bibitem{donagi}
R. Donagi and E. Witten, ``Supersymmetric Yang-Mills 
theory and integrable systems," {\sl Nucl. Phys.} {\bf B460} (1996) 299; 
hep-th/9510101.

\bibitem{gthooft}
G.'t Hooft, ``On the phase transition  towards permanent quark 
confinement," {\sl Nucl. Phys.} {\bf B138} (1978) 1; 
``A property of electric and magnetic flux in non-Abelian gauge theories," 
{\sl Nucl. Phys.} {\bf B153} (1979) 141.

\bibitem{masas}
J.M.F. Labastida and Carlos Lozano, ``Mass perturbations 
in twisted $\cn=4$ supersymmetric gauge theories," {\sl Nucl. Phys.} 
{\bf B518} (1998) 37; hep-th/9711132.

\bibitem{coreatres}
R. Dijkgraaf, J.-S. Park and B.J. Schroers, 
``$\cn=4$ supersymmetric Yang-Mills theory on a K{\"a}hler surface,"
hep-th/9801066.

\bibitem{sun}
J.M.F. Labastida and C. Lozano, to appear.

\bibitem{izzy}
C. Itzykson, ``Level one Kac-Moody characters and modular invariants,"
{\sl Nucl. Phys. B (Proc. Suppl.)} {\bf 5} (1988) 150.
                                                               
\bibitem{estrings} J.A. Minahan, D. Nemeschansky, C. Vafa and N.P. Warner, 
``E-strings and $\cn$=4 topological Yang-Mills theories," {\sl  Nucl. Phys.}
{\bf B527} (1998) 581; hep-th/9802168.

\bibitem{yoshioka}
K. Yoshioka, ``Betti numbers of moduli of stable sheaves on some 
surfaces," {\sl Nucl. Phys. B (Proc. Suppl.)} {\bf 46} (1996) 263; 
``Chamber structure of polarizations and the moduli 
of stable sheaves on a ruled surface," {\sl Int. J. Math.} {\bf 7} (1996) 
411.




\end{thebibliography}
\end{document}